%% file: main.tex
\documentclass[runningheads]{llncs}
\usepackage[T1]{fontenc}
\usepackage{graphicx}
\usepackage[misc]{ifsym}
\usepackage{amsmath}
\usepackage{graphicx}
\usepackage[draft]{hyperref}
\usepackage{booktabs}

\begin{document}
\title{The Impact of Generative Artificial Intelligence on Market Equilibrium: Evidence from a Natural Experiment}
\author{Kaichen Zhang\inst{1} \and
Zixuan Yuan\inst{2} \and
Hui Xiong\inst{1,3}\textsuperscript{(\Letter)}}
\authorrunning{K. Zhang et al.}
\institute{
Thrust of Artificial Intelligence, The Hong Kong University of Science and Technology (Guangzhou), Guangzhou, China \\ \email{kzhangbi@connect.ust.hk}  \and
Thrust of Financial Technology, The Hong Kong University of Science and Technology (Guangzhou), Guangzhou, China \\ \email{zixuanyuan@hkust-gz.edu.cn}  \and
Department of Computer Science and Engineering, The Hong Kong University of Science and Technology, Hong Kong, China \\ \email{xionghui@ust.hk}  
}

\maketitle              % typeset the header of the contribution
\input{sections/0.abstract}
\input{sections/1.introduction}

\input{sections/2.related}
\input{sections/3.strategy}
\input{sections/4.context}

\input{sections/5.findings}

\input{sections/7.demand}

\input{sections/8.supply}

% \input{sections/7.implication}
\input{sections/8.conclusions}

\newpage

\bibliographystyle{splncs04}
\bibliography{reference}

\appendix
\section{Appendix}
\begin{figure}[h]
\centering
\includegraphics[width=0.8\textwidth]{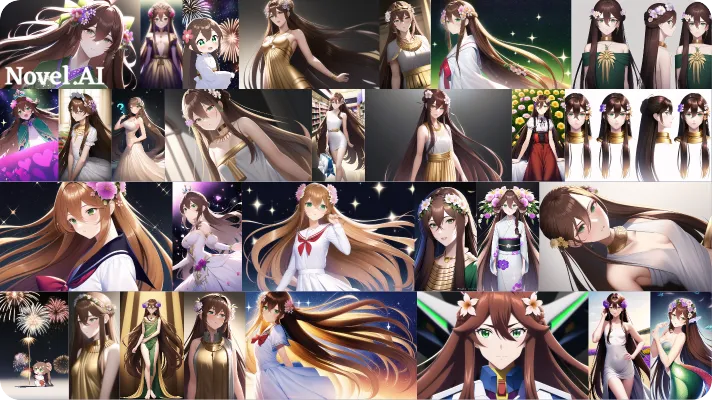}
% \vspace*{-3mm}
\caption{An illustration of the style and quality of NovelAI's AI-generated images.}
\label{fig:novelai}
\end{figure}

\end{document}

%% file: sections/0.abstract.tex
\begin{abstract}
% Generative artificial intelligence (AI) has the capacity to generate creative content like humans, but in much less time and at a much lower cost, raising concerns and questions about its influence on human creators, clientele, and their businesses. 

% Generative artificial intelligence (AI) has the capacity to generate creative content like humans, but in much less time and at a much lower cost, raising concerns and questions about its potential job displacement on human creators. 

Generative artificial intelligence (AI) exhibits the capability to generate creative content akin to human output with greater efficiency and reduced costs. This groundbreaking capability, however, has ignited a debate regarding its potential to displace human creators. In light of these discussions, this paper empirically investigates the impact of generative AI on market equilibrium, in the context of China's leading art outsourcing platform. We overcome the challenge of causal inference by identifying an unanticipated and sudden leak of an advanced image-generative AI as a natural experiment. This leak precipitated a notable reduction in the production costs of anime-style images compared to other genres, thereby providing a unique opportunity for difference-in-differences comparisons. Our analysis shows that the advent of generative AI led to a 64\% reduction in average prices, yet it simultaneously spurred a 121\% increase in order volume and a 56\% increase in overall revenue. This growth is primarily driven by the rising demand for "low-end" personal orders, rather than commercial orders. Moreover, incumbent creators retain the majority of the market share and reap the most benefits of generative AI. 
% The results are further explained theoretically through the lens of market equilibrium theory. 
Our research highlights the potential of generative AI to benefit all stakeholders across the platform economy, yielding both scholarly contributions and practical implications.

\keywords{Artificial Intelligence \and Generative AI \and Market Equilibrium \and Online Platform \and Platform Economy}

\end{abstract}

%% file: sections/1.introduction.tex
\section{Introduction}
Generative artificial intelligence (AI) appears to propel our world closer to the realm of Cyberpunk—a futuristic milieu characterized by the coexistence of "high tech and low life": Generative AI demonstrates its ability to generate text, images, and videos that are similar to content created by human experts – but in much less time, at a fraction of the cost, and with amazing creativity \cite{wesselgenerative}. Such advances, however, have raised significant concerns about the potential for job displacement. One notable manifestation is the organized opposition of artists, who have taken to social media to voice their opposition through campaigns such as "NO to AI-generated images" \cite{NOAI}.

However, the potential impact of generative AI remains controversial in academia. Distinguished AI scientists such as Turing Award winners Yoshua Bengio, Yann LeCun, and Geoffrey Hinton - recognized with what is often called the "Nobel Prize of computer science" - hold divergent views. While Bengio and Hinton have publicly expressed concern, warning of "the risk of extinction by AI," LeCun dismisses such fears as unfounded, citing his understanding of the limitation of current AI systems. This debate echoes the broader discourse found in the economics literature, which suggests that while information technology may carry the risk of rendering certain jobs obsolete \cite{acemoglu2011skills,brynjolfsson2018can,horton2019death,brynjolfsson2019does,yilmaz2023ai}, it may also benefit workers by increasing their productivity \cite{brynjolfsson2019artificial,peng2023impact,noy2023experimental}.

To this end, this paper investigates the impact of generative AI on market equilibrium, with empirical data from China’s leading art outsourcing platform. Specifically, we seek to answer the following research questions: \textbf{RQ1}: In what ways does generative AI shape market equilibrium in terms of average price, order volume, and overall revenue? \textbf{RQ2}: What types of demand are eliminated or created by generative AI? \textbf{RQ3}: Who emerges as the primary suppliers in the market—are they incumbents or entrants utilizing generative AI?

The main challenge in assessing the impact of generative AI lies in achieving \textit{causal inference}. This involves establishing a true cause-and-effect relationship between generative AI and the focal outcomes, rather than mere correlation. One straightforward approach is to contrast the outcome variable (e.g., the number of orders) after and before the emergence of generative AI. However, this method lacks the capacity for causal inference. For example, the demand for anime images could naturally fluctuate due to the seasonal effect of a new semester for students. This intrinsic variability could potentially bias our estimates of the impact of generative AI, leading to incorrect estimates. Another approach is randomized experiments (or A/B testing), which offer the most compelling evidence for causality. However, using experiments to study the impact of generative AI on markets is not feasible. Unlike individuals, markets cannot be randomly assigned to different groups, which limits the ability to implement generative AI effects only on the treated group while leaving the control group unaffected. Macro-level analyses involving countries, cities, and industries face similar limitations.

\begin{figure*}[t]
  \centering
  \includegraphics[width=\textwidth]{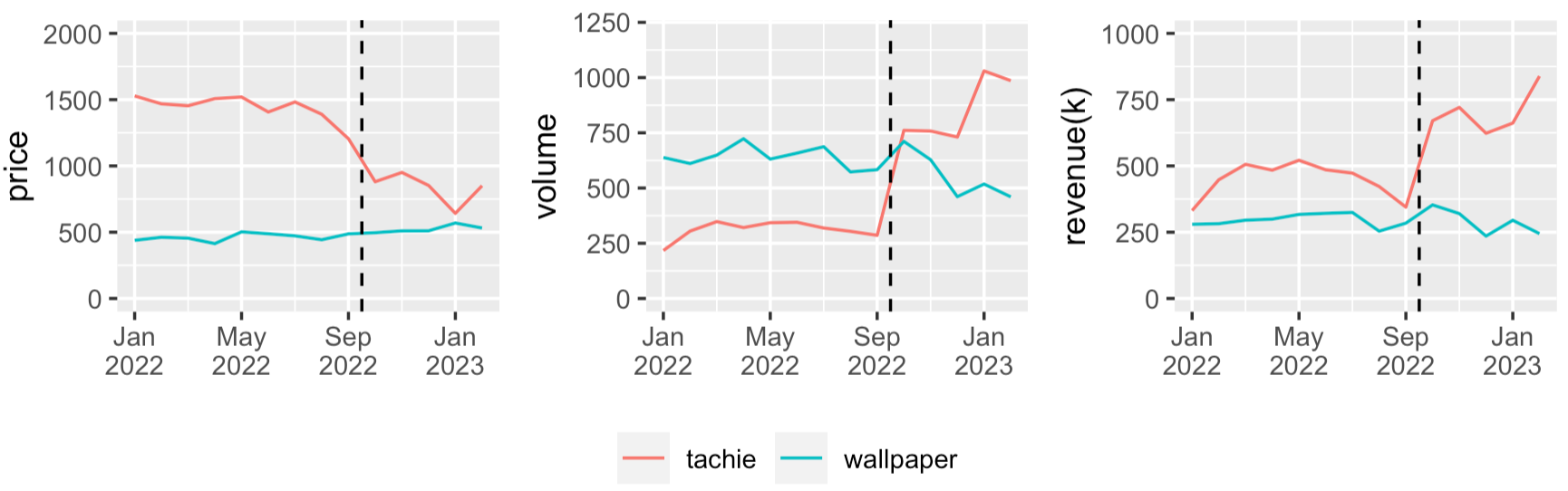}
  \caption{Model-free evidence. The three graphs show the trajectories of average price, order volume, and overall revenue for the "wallpaper" and "tachie" (anime character paintings) markets. The dashed lines represent the leak moment of an advanced anime generative AI, which only affects only the "tachie" market. The post-leak landscape is intriguing: While the "tachie" market experiences a dip in price, it experiences a surge in both volume and revenue, marking a departure from the intuition of job replacement and illustrating the transformative potential of generative AI for market prosperity.}
  % \caption{Model-free evidence. The three graphs show the price, volume, and turnover trends over time for two art categories, "wallpaper" and "tachie," anime character paintings. The dashed lines in three graphs represent the leak of an anime generative AI, which is expected to affect the "tachie" market but not the "wallpaper" market. After the leak, although the price of "tachie" decreased, its volume and turnover increased significantly, suggesting that generative AI lowers price but increases volume and turnover.}
  \label{fig:evidence}
  % \vspace{-5mm}
\end{figure*}

To overcome the challenge of causal inference, we identify an unanticipated and sudden leak of a highly proficient image-generative AI as a \textit{natural experiment}, where markets inadvertently encounter experimental and control conditions, resembling a scenario wherein nature orchestrates an experiment.
% To overcome the challenge of causal inference, we identify a sudden and unexpected leak of a well-trained image-generative AI as a novel \textit{natural experiment}, in which markets are exposed to the experimental and control conditions as if nature or hackers conduct an experiment. 

Among the pioneer platforms in the field, NovelAI stands out as an early entrant, offering text-to-image generative AI services. These services are available for a subscription fee ranging from \$10 to \$25 per month per user. Powered by Stable Diffusion \cite{rombach2022high}, NovelAI's AI can quickly generate anime-style images in response to specific keyword-based requests. However, in early October 2022, hackers attacked NovelAI and leaked the AI model and training codes used by NovelAI. As a result, the leaked resources are available for download and local implementation, effectively reducing the cost of image generation to nearly zero.
% Among the pioneer platforms in the field, Novelai stands out as an early entrant, offering text-to-image generative AI services. These services come at a subscription fee of \$10 to \$25 per month per user. Powered by Stable Diffusion \cite{rombach2022high}, Novelai's AI can swiftly generate anime-style images in response to specific keyword-based requests. However, in early October 2022, hackers attacked Novelai and leaked the AI model and training codes used by Novelai. Consequently, the leaked resources are accessible for download and local implementation, effectively reducing the cost of image generation to nearly zero.
% Novelai is one of the earliest websites, which provide text-to-image generative AI services, charging \$10 to \$25 per month for a subscription for each user. Its AI, based on Stable Diffusion \cite{rombach2022high}, can generate anime-style images in a few seconds, at the specified requests by keywords. However, at the beginning of October 2022, hackers attacked Novelai and released Novelai's AI model and training codes. As a consequence, anyone can download the leaked material and implement the same AI locally, which reduces the cost of generating an image to almost zero. 

The inadvertent leak, which originated in the Western world, has inadvertently reverberated throughout the Eastern world. Prior to October 2022, China lacked domestic providers of generative AI, while access to foreign alternatives was also impeded. After October 2022, however, the leaked AI, along with its accompanying tutorials, quickly spread throughout China's social media landscape. As a result, this leak has expeditiously transitioned China from a state of being unaffected by generative AI to one of being affected by it.
% The inadvertent leak, which originated in the Western world, has inadvertently reverberated throughout the Eastern world. Prior to October 2022, China lacked domestic providers of generative AI, while access to foreign alternatives was also impeded. After October 2022, however, the leaked AI, with accompanying tutorials, rapidly spread throughout China's social media landscape. Consequently, this leak has expeditiously transitioned China from a state of being unaffected by generative AI to one of being affected.
% This leak that happened in the Western world inadvertently affects the Eastern world. Before October 2022, there is no local generative AI provider in China and people can not access foreign generative AI providers because of policy restrictions, language restrictions, and payment restrictions. The awareness of generative AI was only limited to researchers in China. However, after October 2022, the leaked AI, along with related tutorials, was widely spread in China's social media. As a result, this leak has taken China from not being affected by generative AI to being affected by generative AI, in a very short period of time.

This leak served as an ideal natural experiment. First, China constitutes an environment where painting markets remained unaffected by generative AI prior to the leak due to policy restrictions, language restrictions, and payment restrictions. Second, the leaked AI  specializes exclusively in generating anime-style images, thus affecting only the anime painting markets while leaving other painting styles unaffected, allowing for a comparison between treated and control units. Third, the unanticipated and sudden nature of the leak is exogenous. This feature ensures that the focal markets are unlikely to have been affected by unobserved confounding factors as stakeholders did not have time to react to the leak in advance. Thus, alternative explanations are effectively ruled out.

We collected 197,068 records from a leading art outsourcing platform in China from January 2022 to February 2023. This platform allows consumers to place orders with specifications and budgets, and artists as freelancers to fulfill them. The orders are segmented into distinct markets, such as anime, UI, live2d, and wallpaper. We transformed these order records into a market-week panel dataset, focusing on average price, order volume, and overall revenue. Using a difference-in-differences research design, we compare the anime market to the wallpaper market, before and after the unintended leak of NovelAI's anime generative AI. Our main findings are summarized in Figure~\ref{fig:evidence}.
% We collected 197,261 records from January 2022 to March 2023 from one of the largest painting outsourcing platforms in China, where consumers can post an order with specified prices and requirements, and artists can undertake those orders. Orders are categorized into different markets, such as anime, UI, live2d, and wallpaper markets. We aggregate the records into a market-week panel dataset, which contains the average price, number of orders, and total revenues as our focal outcome variables. We adopt a difference-in-differences (DID) design to identify the impact of generative AI, in which we exploit the leak as a natural experiment and compare the focal outcome variables between the anime market as the treated unit and the wallpaper market as the control unit. Figure~\ref{fig:evidence} shows the model-free evidence of our findings. 

Our analysis reveals a promising effect on market equilibrium following the emergence of generative AI. Specifically, although this emergence leads to a 64\% reduction in the average price, the generative AI also leads to a remarkable 121\% increase in order volume and a consequent 56\% increase in overall revenue. All effects are statistically significant at the 1\% level. In addition, the effects are also economically significant, due to their substantial magnitude. The results suggest that generative AI can contribute to a more prosperous market, where each stakeholder benefits: consumers get cheaper services, and human creators and the platform get more orders and revenue.

% We find that the emergence of generative AI decreases the transaction price but surprisingly increases the transaction volume and turnover. In particular, though the average price decreases by XXX\%, the number of transactions increases by XXX\%, and the total transaction revenues increase by XXX\%. All the estimates are statistically significant. This finding shows that instead of putting artists out of work, generative AI has increased their income and made the industry more prosperous. 

% We explain and generalize our findings through the lens of market equilibrium theory. On the one hand, the emergence of generative AI has the potential to reduce demand, as consumers gain the ability to satisfy their needs autonomously using this technology. On the other hand, generative AI simultaneously increases supply as producers become more efficient in their production. These two opposing dynamics can lead to a decrease in prices and an increase in order volume and revenue, which explains our observed empirical results.

% We provide a theoretical explanation for this counter-intuitive finding through the supply-and-demand model: on the one hand, generative AI will decrease the demand, because consumers can fulfill their demands by themselves using generative AI. But on the other hand, generative AI also increases supply, because producers become more efficient. Two counteracting forces together can lead to our empirical finding.

We then examine what types of demand are created by generative AI. Orders on the platform are categorized into personal orders and commercial orders, with different licenses. We conduct a heterogeneous analysis, showing that the growth of generative AI is mainly driven by personal orders, rather than commercial orders. The "tachie" market achieved significantly more revenue through more personal orders at low prices, while the price, volume, and revenue of commercial orders are relatively stable. Our results indicate the potential of generative AI for business success by satisfying the "low-end" needs of individual users.
% We then examine what types of demand are created by generative AI. The orders on the focal platform are categorized into personal orders and commercial orders, with distinct licenses of copyrights. We conduct a heterogeneous effect, showing that the growth of generative AI is mainly attributed to personal orders, rather than commercial orders. The "tachie" market achieved remarkably more revenue through more personal orders at low prices, while the price, volume, and revenue of commercial orders are relatively stable. Our results indicate the potential of generative AI for business success by satisfying the "low-end" needs of individual users.

Finally, we examine who emerges as the primary suppliers in the market. Our results show that the incumbents, who registered before the leak, won 97\% of the orders after the leak. As a result, they reap most of the benefits of increased order volume and revenue. The results rule out the possibility that new entrants using generative AI have taken over the market share of real human creators.

Our work provides valuable managerial implications. First, the increased order volume and overall revenue indicate more jobs and taxes, providing favorable economic evidence of generative AI for policymakers to make supportive AI policies. Second, generative AI mainly benefits incumbent creators, suggesting that they should welcome and incorporate generative AI into their workflow to further increase their productivity. Third, online platforms can conduct marketing campaigns and improve services specifically for individual users to better exploit the blue ocean driven by generative AI.
% The paper's findings have significant societal implications. Practitioners should integrate generative AI into their workflows rather than resist it. Embracing this technology can increase efficiency, ultimately translating to increased revenue. Policymakers, on the other hand, should view the increase in total revenue as a broader tax base. In this light, encouraging the utilization of generative AI becomes a more favorable approach than imposing restrictions. For AI communities, in addition to advocating for greater investment, this study could assuage any unease and guilt that researchers may feel about the potential societal upheaval caused by AI.
% The finding in this paper has great social implications. For practitioners, they should incorporate generative AI as a part of their working flows, instead of resisting it, because generative AI can increase their revenue by improving their efficiency. For policymakers, a higher total revenue indicates a larger base for taxes, which indicates a promotion instead of a restriction of generative AI. For AI communities, this paper provides cited evidence that generative AI has benefited society. Besides this paper may encourage researchers by eliminating their inside guilty of "AI destroying the world".

In conclusion, this paper makes the following contributions:
\begin{itemize}
\item To the best of our knowledge, we pioneer the estimation of the impact of generative AI on market equilibrium, making scholarly contributions to the growing literature on the economics of (generative) AI. 
\item We provide promising empirical evidence that generative AI can promote market prosperity and explain the reason. We also provide valuable managerial insights for various stakeholders.
\item Our identified natural experiment facilitates causal inference for future research on generative AI—its effects on customer satisfaction, on employment, and beyond.

% \item This paper pioneers the estimation of the impact of generative AI, by exploiting an unexpected leak of a commercial generative AI as a novel natural experiment.
% \item This paper presents a counterintuitive finding that generative AI does not replace practitioners but instead contributes to industry prosperity, with significant societal implications for employment, legislation, and AI development.
% \item The natural experiment identified in this paper establishes a foundation for further research on the causal effects of generative AI—its effects on quality in product markets, on wages in labor markets, and beyond.

    % \item This paper among the first estimates the impact of generative AI, by exploiting a sudden and unexpected leak of a commercial generative AI as a novel natural experiment.
    % \item This paper shows a counter-intuitive finding that generative AI increases instead of decreases volume and turnover in a product market. This finding has great social implications for employment, legislation, and the development of AI.
    % \item The natural experiment identified in this paper enables follow-up work on the causal impact of generative AI, which includes but is not limited to the impact on other dependent variables in product markets, such as quality, or the impact in other areas, such as salary in labor markets.
\end{itemize}

%% file: sections/2.related.tex
\section{Related Literature}
This paper adopts empirical research methods to study generative AI, which is related to the field of generative AI and the economics of AI.
\subsection{Generative AI}
Generative AI is defined as artificial intelligence (AI) algorithms that generate original outputs based on prompt inputs, which is distinguished from discriminative AI in that generative AI synthesizes the data it has been trained on and creates similar content created by human experts \cite{wesselgenerative}.

Generative AI has many applications, which their output format can categorize. For text, the applications include machine translation \cite{bahdanau2014neural}, code generation \cite{svyatkovskiy2020intellicode}, natural language generation \cite{gatt2018survey}, etc. For images, the applications include text-to-image generation \cite{ramesh2021zero}, style transfer \cite{gatys2016image}, image translation \cite{isola2017image}, etc. For audio, the applications include text-to-speech \cite{ren2019fastspeech}, music composition \cite{fernandez2013ai}, singing voice conversion \cite{mohammadi2017overview}, etc. 

Generative AI's underlying generative models include Autoregressive Model (AR) \cite{model2013system}, Generative Adversarial Network (GAN) \cite{goodfellow2014generative}, Normalizing Flow (Flow) \cite{rezende2015variational}, Variational Auto-Encoder (VAE) \cite{kingma2013auto}, and Denoising Diffusion Probabilistic Model (Diffusion) \cite{ho2020denoising}. For example, OpenAI's ChatGPT adopts an autoregressive model \cite{brown2020language} with Transformer architecture \cite{vaswani2017attention}. The focal generative AI in this paper, NovelAI's model, adopts stable diffusion \cite{rombach2022high}.

Stable diffusion is a cutting-edge text-to-image deep learning model that generates images based on textual descriptions. Training Stable Diffusion takes two steps: initially, a large dataset of text-image pairs is used to teach the model the complex relationships between textual descriptions and visual content; subsequently, the model learns to iteratively refine images from noise, given these text inputs.

The technical foundations of stable diffusion are relevant to our work in the following ways. First, stable diffusion can only generate images from the same distribution of its training set. Since NovelAI's model is trained on anime images, it can only generate anime images, allowing for comparison across genres. Second, stable diffusion may struggle with maintaining fidelity to specific details in the text prompts, especially for complex or highly detailed descriptions, resulting in images that are somewhat generic or miss nuanced aspects of the desired output. This limitation can qualitatively explain why generative AI mainly affects low-end orders and cannot replace incumbent artists.

\subsection{The Economics of AI}
The economics of artificial intelligence (AI) refers to the study of how AI affects markets, organizations, and individuals, and how it changes the way businesses operate and make strategic decisions.

The main research method for studying the economics of AI is empirical research, which meticulously examines the impact of AI adoption on a focal area through the careful analysis of observational data and controlled experiments. Past research has examined the impact of AI on areas such as innovation \cite{allam2016impact}, human resource management \cite{tambe2019artificial}, sales \cite{luo2021artificial}, consumer behavior \cite{sun2019effect}, international trade \cite{goldfarb2019artificial}, education \cite{popenici2017exploring} and healthcare \cite{wolff2020economic}.

A stream of papers on the impact of AI on the labor market \cite{webb2019impact,frank2019toward} is closely related to our work. These papers mainly focus on how AI affects the employment and wages of workers, and who is more or less likely to be affected by AI. The literature reveals two contradictory aspects: on the one hand, AI may carry the risk of job displacement \cite{acemoglu2011skills,brynjolfsson2018can,horton2019death,brynjolfsson2019does,yilmaz2023ai}; on the other hand, AI may also benefit workers by increasing their productivity \cite{brynjolfsson2019artificial,peng2023impact,noy2023experimental}.

% The main research method for studying the economics of IT is empirical studies of observational data and experiments. Research questions usually take the form of what the impact of X on Y is. For example, the authors \cite{pamuru2021impact} examine the impact of an augmented reality game on local businesses, i.e., Pokémon Go on restaurants. In the context of AI, the research includes the impact of AI on education \cite{popenici2017exploring}, labor \cite{frank2019toward}, healthcare \cite{wolff2020economic}, etc.

% The vision paper \cite{brynjolfsson2021economics} summarizes three criteria for evaluating new economics of IT research: questions should be novel and interesting; methods should be correct and useful; outcomes should be important and impactful.

Our paper contributes to the literature on the economics of (generative) artificial intelligence. First, we pioneer the estimation of the impact of generative AI, the new era of AI whose effects are still unclear and understudied. Second, our research target, a marketplace, is unique and provides perspectives from both the supplier and consumer sides, providing more insights into the impact of generative AI. Third, our identified natural experiment facilitates future research on generative AI by identifying an ideal scenario for causal inference.

% This paper squarely meets the outlined criteria for several reasons. Firstly, it pioneers the exploration of generative AI's impact, a field that has rapidly gained significant societal attention since its inception. Secondly, it deftly navigates the challenge of causal inference through the ingenious identification of a novel natural experiment, which also establishes a foundation for future research. Finally, the paper's result is unexpected in that what was thought to be negative is actually positive, which has important implications for various stakeholders.
% This paper fits the criteria in that: this paper is among the first to study the impact of generative AI, which has created a great social impact since it was just invented; this paper overcomes the challenge of causal inference by identifying a novel natural experiment, which can also be used for other researches; this paper's finding is unexpected as what was thought to be negative is really positive, which has great implications for various stakeholders.

%% file: sections/3.strategy.tex
\section{Empirical Strategy}
In this section, we compare different empirical methods for estimating the causal effects of generative AI. We apply our analysis under the potential outcome framework \cite{rubin2005causal}, the most widely used framework for causal inference.

\subsection{Pretest-Posttest Design and Experimental Design}
Let us consider the case of analyzing a set of $N$ units, labeled from $1$ to $N$. For each unit $i$, there are two potential outcomes $\{Y_{1i},Y_{0i}\}$ depending on whether the unit receives treatment or not ($D_i=\{1,0\}$). The individual treatment effect for unit $i$ is then defined as $Y_{1i}-Y_{0i}$. In this paper, the units are markets of different art genres, the treatment is the emergence of generative AI and the outcomes are market equilibrium variables, i.e., price, volume, and revenue.

The main challenge in making causal inferences is posed by the fact that there is only one potential outcome that can be observed for each unit. For treated units ($D_i=1$), we can only observe $Y_{1i}$; for control units($D_i=0$), we can only observe $Y_{0i}$. As a result, individual treatment effects can never be estimated unless strong assumptions about the data generation process are made. As an alternative, researchers estimate the average treatment effect (ATE, $E(Y_{1i}-Y_{0i})$) or the average treatment effect on the treated (ATT, $E(Y_{1i}-Y_{0i}|D_i=1)$).

We further extend our demonstration to two time periods. In the pretest period ($t=0$), all units are not treated, where we can observe $E(Y_0(t=0)|D=0)$ and $E(Y_0(t=0)|D=1)$. In the posttest period ($t=1$), the treated units ($D=1$) have received the treatment while the control units ($D=0$) remain untreated. We can observe $E(Y_0(t=1)|D=0)$ and $E(Y_1(t=1)|D=1)$. Our goal is to estimate ATE ($E(Y_1(1))-E(Y_0(1))$) or ATT ($E(Y_1(1)|D=1)-E(Y_0(1)|D=1)$). Note that we still cannot observe another potential outcome ($E(Y_0(1)|D=1)$ and $E(Y_1(1)|D=0)$).

A straightforward yet unsophisticated method is the pretest-posttest design, which compares the outcome variable before and after the treatment. Its estimator is $E(Y_1(1)|D=1)-E(Y_0(0)|D=1)$. However, ATT equals $E(Y_1(1)|D=1)-E(Y_0(1)|D=1)=(E(Y_1(1)|D=1)-E(Y_0(0)|D=1))-(E(Y_0(1)|D=1)-E(Y_0(0)|D=1))$, which indicates that the estimator of the pretest-posttest design is unbiased only if $E(Y_0(1)|D=1)-E(Y_0(0)|D=1)=0$, i.e., the outcome variable does not change for the treated units if they do not receive the treatment.

However, the equation does not hold in the context of this study. A plausible explanation for this discrepancy could be the influence of seasonal effects. For example, the demand for anime images could naturally fluctuate after October 2022 due to the start of a new academic semester for students. This intrinsic variability could potentially skew our estimates of the impact of generative AI, leading to underestimates, overestimates, or even fundamentally incorrect estimates. As a result, it's clear that the pretest-posttest design is not an appropriate approach for this study.

% However, the equation is unlikely to hold in this study. One possible reason is the presence of seasonal effects.
% For example, the demand for anime paintings may increase/decrease after October 2022 due to the start of a new semester for students, which could lead to either an underestimation, overestimation or a fundamentally incorrect estimate of the impact of generative AI. As a result, the pretest-posttest design is not appropriate for this study.

Another method is randomized experiments, also called a/b tests in some literature. This method randomly assigns units to a treated group ($D=1$) and a control group ($D=0$), ensuring that the population of the treated group is the same as the population of the control group. So we have $E(Y_0(1)|D=1)=E(Y_0(1)|D=0)$. We can then derive ATT by subtracting $E(Y_1(1)|D=1)$ from $E(Y_0(1)|D=0)$, since we can observe both parts. Also, ATT is equal to ATE in randomized experiments. Thus, randomized experiments are considered the best method for causal inference.

However, randomized experiments are not feasible in this study. While the individual level of analysis allows for the division of subjects into treatment and control groups, the challenge arises at the market level of analysis. Unlike individuals, markets cannot be randomly assigned to different groups, which limits the ability to implement generative AI effects only on the treated group while leaving the control group unaffected. Macro-level analyses involving countries, cities, and industries face similar limitations. In these contexts, researchers have to gain valuable insights from observational data rather than from controlled interventions.
% Despite its power, we cannot conduct random experiments for this study. When the level of analysis is individuals, we can divide them into treated and control groups. However, the level of analysis in this study is markets. We cannot randomly assign markets to different groups and apply the effects of generative AI only to the treated group but not to the control group. For the macro level of analysis, such as countries, cities, and industries, researchers can hardly conduct random experiments and can only get their results from observational data.

\subsection{Difference-In-Differences Design}
Difference-in-differences (DID) is arguably the most popular research design in the quantitative social sciences. It can be traced back to 1855 \cite{snow1856mode} and has also made great strides in recent years \cite{goodman2021difference}.

The DID estimator is $E(Y_1(1)|D=1)-E(Y_0(0)|D=1)-(E(Y_0(1)|D=0)-E(Y_0(0)|D=0))$. DID first calculates the differences between the pretest and posttest periods for each group. DID then takes the difference between these two differences as the estimator. 

The DID design is based on the parallel trend assumption that the trends in the outcome variable for the treatment and control groups would have been parallel in the absence of the treatment. Formally, the DID estimator equals ATT if $E(Y_0(1)|D=1)-E(Y_0(0)|D=1)=E(Y_0(1)|D=0)-E(Y_0(0)|D=0)$. The parallel trend assumption can be deemed satisfied if the treatment and control groups exhibit comparable pre-treatment trends in the outcome variable.

The application of the difference-in-differences to examine the impact of generative AI on market equilibrium hinges on three pivotal conditions: an appropriate treatment that affects treated markets subsequent to the treatment, not prior to it; minimal impact on control markets; and the fulfillment of the parallel trend assumption between treated and control markets.

% In summary, to apply the DID to study the impact of generative AI on product markets, we should find a treatment of generative AI such that it affects treated markets after the treatment and not before the treatment; it does not affect control markets; treated and control markets satisfy the parallel trend assumption.

This study employs the difference-in-differences design, utilizing an unanticipated leak of an advanced generative AI as the treatment. Prior to the leak, the trends of the treated and control markets were stable and parallel, as illustrated in Figure~\ref{fig:evidence}, which lends support to the validity of the parallel trend assumption. 

% We further conduct two statistical tests to verify this assumption in Appendix \ref{app:parallel}.

% This study employs the difference-in-differences design, utilizing an unanticipated leak of an advanced generative AI as the treatment. The parallel trends of the treated and control markets, as illustrated in Figure 1, serve to substantiate the efficacy of the DID design. This alignment of outcome variables prior to the leak, characterized by stability, lends support to the validity of the parallel trend assumption.

% This paper adopts the DID design by exploiting a sudden and unexpected leak of a well-trained image-generative AI as a natural experiment. As shown in Figure~\ref{fig:evidence}, the treated and control unit followed the same trend before the leak, where the outcome variables are stable before the leak, which verifies our DID design by showing that the parallel trend assumption holds.

%% file: sections/4.context.tex
% \begin{figure}[!]
% \centering
% \includegraphics[width=0.45\textwidth]{images/novelai.png}
% % \vspace*{-3mm}
% \caption{An illustration of the style and quality of NovelAI's AI-generated images.}
% \label{fig:novelai}
% \vspace*{-3mm}
% \end{figure}

\section{Research Context and Data}
\subsection{The NovelAI "experiment"}
NovelAI is one of the first commercial websites offering generative AI services. NovelAI was initially launched on June 15, 2021, when it only offered AI story-writing services. After Stable Diffusion, a state-of-the-art text-to-image algorithm, was released on March 2, 2022, NovelAI began offering its text-to-image services. NovelAI put a lot of effort into collecting anime-style images and used them to train an anime-style generative AI model based on Stable Diffusion. NovelAI's text-to-image service was implemented on October 3, 2022 \cite{novelai2022}.

NovelAI's generative AI takes a series of prompts, or keywords, such as "masterpiece, magical girl, fantasy, caustic" as input. The generative AI will generate images according to the prompts. 
Figure~\ref{fig:novelai} in the Appendix presents the official examples of NovelAI. 
Note that NovelAI's AI only generates anime-style images because it is trained on anime-style images.

NovelAI's generative AI can generate an image in a few seconds and generate multiple images simultaneously. The cost of generating an image depends on its quality and resolution size. NovelAI uses a monthly subscription scheme ranging from \$10 to \$25. \$10 users can generate approximately 200 images of normal quality per month. \$25 users can generate an unlimited number of images.

However, on October 6, 2022, a few days after NovelAI's text-to-image service was released, hackers attacked NovelAI's server and released its AI model and associated training code. Anyone can download the leaked material via a shared seed and implement the same AI locally, allowing people to generate an image in a few seconds for free.

\begin{figure}[t!]
\centering
\includegraphics[width=1\textwidth]{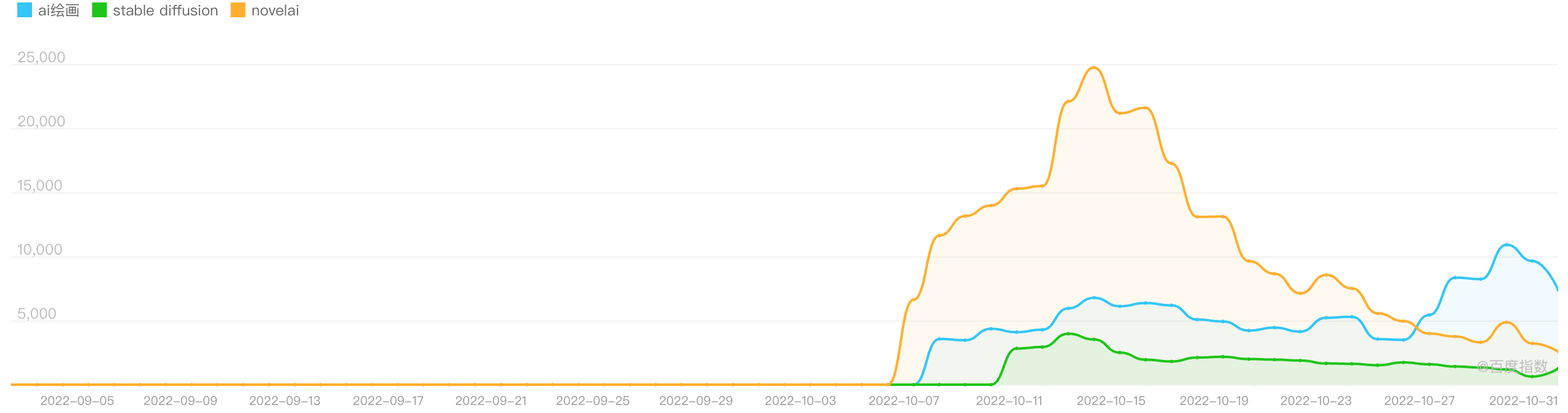}
% \vspace*{-3mm}
\caption{Search Engine Trend of AI Painting (Blue), Stable Diffusion (Green), and NovelAI (Orange).}
\label{fig:trend}
\vspace*{-3mm}
\end{figure}

This leak served as an ideal natural experiment for several reasons. First, China constitutes an environment where painting markets remained unaffected by generative AI before the leak. China did not yet have a local generative AI provider, and people cannot access foreign generative AI providers due to policy restrictions, language restrictions, and payment restrictions. Figure~\ref{fig:trend} shows the search trend from Baidu, the largest search engine in China. The trend verifies that painting markets had not been affected by generative AI before the leak. Second, the leaked AI exclusively specializes in generating anime-style images, thereby exerting an impact solely on the anime painting markets while leaving other painting styles unaffected, allowing for a comparison between treated and control units. Third, the leak caused by hackers is unanticipated and sudden, which contributes to exogeneity. This attribute ensures that the focal markets are unlikely to have been affected by unobserved confounding factors since stakeholders had no time to react to the leak in advance. As a result, alternative explanations beyond the emergence of generative AI are effectively ruled out.

% This leak creates an ideal natural experiment in China. First, China provides an environment where painting markets had not been affected by generative AI before the leak. China did not yet have a local generative AI provider, and people cannot access foreign generative AI providers due to policy restrictions, language restrictions, and payment restrictions. However, since October 2022, the leaked AI and related tutorials have been widely disseminated on Chinese social media, as if the leak had created an experimental treatment in China. Second, the leaked AI only generates anime-style images. Thus, it only affects the anime painting markets, but not the painting markets of other styles, which allows a comparison between the treated units and the control units. Third, the leak was sudden and unexpected, creating an exogenous natural experiment. The product market was unlikely to be affected by unobserved confounders, as stakeholders cannot react prior to the leak, ruling out alternative explanations other than generative AI.

\subsection{Data Description}
We collect data from Mihuashi, the leading art outsourcing platform in China. Consumers, as demanders, can post an order on the focal platform, and artists, as producers, can express their interest in accepting the order. The consumer can then select one of the interested artists and enter into a contract with the selected artist. Once the artist has completed the task, the consumer makes the payment. The platform charges a commission of 5\% of the price of the order.

An order on the focal platform is comprised of an art category, a price range, and a description. The art category includes "original character", "artwork", "avatar", "Q style", "tachie", "wallpaper", "live2d", "scenario", "cover", "poster", "UI", "character design", and "CG". The price range is defined by a lower and an upper bound, such as 50 to 200 yuan. The order description specifies the requirement, such as "a wallpaper for Vtuber live streaming background".

We collected data on the focal platform from January 2022 to February 2023, which contains 197,068 orders in total. We removed 408 outlier orders whose price is more than 70,000 yuan (about \$10,000) because they are too expensive to become real orders. We then aggregate the data to the category-week level. For each category and each week, we calculate the average price (price), order volume (volume), and overall revenue (revenue). We calculate the average price and overall revenue based on the lower bound because the lower bound may better reflect the true willingness of consumers to accept their orders than the upper bound. 

% Our results are robust with alternative price measurements or longer time periods, as demonstrated in Appendix~\ref{app:price} and \ref{app:period}.

\urldef{\urlA}\url{https://en.wiktionary.org/wiki/%E7%AB%8B%E3%81%A1%E7%B5%B5#Japanese}

\input{sections/summary_table}

We select the category "tachie" as the treated market. The "tachie" is a Japanese term in the context of anime, manga, and video games that refers to the full-body standing drawing of a character.\footnote{\urlA} The "tachie" is precisely the style that NovelAI's AI produces. In contrast, we select the "wallpaper" as the control market. A control unit should be comparable to the treated unit; thus, "live2d," "UI," and "CG" were excluded from consideration because they are beyond the scope of paintings. With regard to the remaining categories, they are likely to be partially affected by NovelAI's generative AI, because the generative AI may only be useful with careful prompt engineering and post-processing, which could result in an underestimation of the effects. The "wallpaper" may be the best control unit because it is unlikely that NovelAI's generative AI will generate elaborate wallpapers. 

Table~\ref{table:summary} presents the summary statistics for the category-week panel dataset in this paper. \textit{Price} refers to the average price. \textit{Volume} refers to order volume. \textit{Revenue} refers to the overall revenue. The currency of \textit{Price} and \textit{Revenue} is Chinese yuan (7 yuan = \$1). Also, we show the price and revenue based on the upper bound, which is \textit{Price\_upper} and \textit{Revenue\_upper}. Panel A presents the statistical data for all orders, while Panel B displays the statistical data for only completed orders. 

% The dataset utilized in this paper is comprised of all orders, and the findings presented herein are consistent when analyzed using completed order data from Appendix~\ref{app:complete}.

% Table~\ref{table:summary} presents the summary statistics for the category-week panel dataset in this paper. Panel A shows the statistics for the "tachie" and "wallpaper" categories; Panel B shows the statistics for all categories. \textit{Price} refers to the average price. \textit{Volume} refers to order volume. \textit{Revenue} refers to the overall revenue. The currency of \textit{Price} and \textit{Revenue} is Chinese yuan (7 yuan = \$1). Also, we show the price and revenue based on the upper bound, which is \textit{Price\_upper} and \textit{Revenue\_upper}.

%% file: sections/summary_table.tex
\begin{table}[t!]
% \linespread{1.5}
% \renewcommand{\arraystretch}{1.5}
\selectfont
\centering
% \resizebox{\linewidth}{!}{
\begin{tabular}[t]{lccccc}
\toprule
  & Mean & SD & Min & Median & Max\\
\midrule
\multicolumn{6}{c}{Panel A: All Orders} \\
% \textit{Price} & 856.8 & 479.4 & 334.6 & 628.0 & 2473.2\\
% \textit{Volume} & 124.1 & 55.0 & 12.0 & 130.0 & 324.0\\
% \textit{Turnover} & 93595.5 & 45687.8 & 6000.0 & 80061.0 & 264504.0\\
% \textit{Price\_upper} & 2323.4 & 934.9 & 1212.2 & 1983.5 & 5175.4\\
% \textit{Turnover\_upper} & 261536.1 & 107168.7 & 21550.0 & 236897.0 & 655574.0\\
\textit{Price} & 843.7 & 460.7 & 334.6 & 628.0 & 2473.2\\
\textit{Volume} & 123.7 & 54.5 & 12.0 & 130.0 & 318.0\\
\textit{Revenue} & 92362.1 & 45096.0 & 6000.0 & 78950.0 & 264104.0\\
\textit{Price\_upper} & 2237.5 & 869.3 & 1212.2 & 1949.5 & 4689.1\\
\textit{Revenue\_upper} & 251996.5 & 103591.0 & 21550.0 & 230810.0 & 656074.0\\

\multicolumn{6}{c}{Panel B: Completed Orders} \\
% \textit{Price} & 852.9 & 890.0 & 0.0 & 567.6 & 10850.0\\
% \textit{Volume} & 240.9 & 377.2 & 0.0 & 87.0 & 2666.0\\
% \textit{Turnover} & 109112.6 & 121770.2 & 0.0 & 64540.0 & 532107.0\\
% \textit{Price\_upper} & 2225.3 & 1824.2 & 0.0 & 1771.0 & 17275.0\\
% \textit{Turnover\_upper} & 402319.9 & 570699.9 & 0.0 & 177300.0 & 3217916.0\\
% \textit{Price} & 872.7 & 806.5 & 50.0 & 599.6 & 7338.7\\
% \textit{Volume} & 252.1 & 381.2 & 1.0 & 94.5 & 2666.0\\
% \textit{Revenue} & 113144.7 & 121162.4 & 50.0 & 66347.5 & 532007.0\\
% \textit{Price\_upper} & 2231.8 & 1614.5 & 100.0 & 1813.6 & 12462.5\\
% \textit{Revenue\_upper} & 402957.1 & 547601.9 & 100.0 & 185405.0 & 3116474.0\\
\textit{Price} & 885.9 & 472.3 & 327.9 & 721.5 & 2321.6\\
\textit{Volume} & 43.2 & 18.1 & 1.0 & 43.0 & 89.0\\
\textit{Revenue} & 	34718.8 & 19238.0 & 500.0 & 30501.0 & 111170.0\\
\textit{Price\_upper} & 2266.1 & 901.7 & 1152.0 & 2010.9 & 5154.8\\
\textit{Revenue\_upper} & 92299.1 & 46196.5 & 1500.0 & 81650.0 & 286370.0\\
\bottomrule
\end{tabular}
% }
\vspace{+3mm}
\caption{Summary Statistics}
\label{table:summary}
\end{table}

%% file: sections/5.findings.tex
\section{Main Analysis} \label{sec:main}
In this section, we answer the research question RQ1, by examining the impact of generative AI on market equilibrium in terms of average price, order volume, and overall revenue.

We employ the standard multi-periods model specification of difference-in-differences with two-way fixed effects \cite{callaway2021difference}:
\begin{equation} \label{equation:main}
y_{it}=\alpha_i +\delta_t + \beta\cdot treated_i after_t + \varepsilon_{it}
\end{equation}

In this model, the variable $y_{it}$ represents one of our outcome variables by category $i$ in week $t$. $treated_i$ is a dummy variable indicating whether category $i$ belongs to the treated group ($treated_i=1$) or the control group ($treated_i=0$). $after_t$ is also a dummy variable indicating whether week $t$ is after the natural experiment ($after_t=1$) or before the natural experiment ($after_t=0$). The model includes a fixed effect $\delta_t$ for each week to control for observed and unobserved nonparametric time trends. Additionally, the model includes a fixed effect $\alpha_i$ for each category to account for time-invariant differences between categories. The parameter of interest is $\beta$, which captures the impact of generative AI on the focal outcome variable.

\input{sections/reg_table}

Table~\ref{tab:reg_main} shows the estimates for the change in price, volume, and revenue in columns (1) to (3), respectively. The coefficient estimate of price is -646, indicating generative AI lowered the average price by 646 yuan. Notably, the coefficient estimates of volume and revenue are 140 and 56,674, showing that generative AI increased the number of orders by 140 per week and increased the total revenue by 56,674 yuan per week. In addition, columns (4) to (6) take the logarithm of price, volume, and revenue and show their percentage change. The results show that price decreased by 64\%, volume increased by 121\%, and revenue increased by 56\%. All effects are statistically significant at the 1\% level.

Public concern about generative AI is often rooted in negative anticipation of the economic ramifications associated with its adoption. In particular, there is an understandable concern among various segments, including artists, that generative AI could lead to feelings of displacement, given the fear of rapid and unceremonious replacement after years of dedicated study and creative toil. 

However, our empirical findings provide a counterintuitive insight, showing that contrary to initial impressions, generative AI can benefit all stakeholders across the platform economy. Consumers can fulfill their orders 64\% cheaper than before, indicating an improvement in consumer welfare. Furthermore, artists receive 56\% more revenue, as each monetary transaction completed by consumers is considered revenue for the artist. They receive 121\% more orders, which underscores the increased opportunities that accelerate the evolution of their artistic endeavors by providing an invaluable platform for rapid prototyping and iterative refinement. Moreover, the platform's revenue also increases, as their commission is proportional to the total revenue. The substantial increase in order volume could further benefit the platform economy through the network effect \cite{uzzi1996sources,mcintyre2017networks}.

%% file: sections/reg_table.tex
\begin{table*}[t]
% \linespread{1}
% \huge
% \selectfont
% \footnotesize
\centering
    % \vspace{-5mm}
    \resizebox{\linewidth}{!}{
   \begin{tabular}{lcccccc}
      \tabularnewline \midrule \midrule
      Dependent Variables:    & price                   & volume                  & revenue                 & log(price)              & log(volume)             & log(revenue)\\  
      Model:                  & (1)                     & (2)                     & (3)                     & (4)                     & (5)                     & (6)\\  
      \midrule
      \emph{Variables}\\
      treated $\times$ after  & -646.5$^{***}$          & 140.3$^{***}$           & 56,674.4$^{***}$        & -0.643$^{***}$           & 1.21$^{***}$             & 0.567$^{***}$\\   
                              & ($2.0\times 10^{-12}$)  & ($4.9\times 10^{-13}$)  & ($1.7\times 10^{-10}$)  & ($1.8\times 10^{-15}$)  & ($3.8\times 10^{-15}$)  & ($1.5\times 10^{-15}$)\\    
      \midrule
      \emph{Fixed-effects}\\
      category           & Yes                     & Yes                     & Yes                     & Yes                     & Yes                     & Yes\\  
      week              & Yes                     & Yes                     & Yes                     & Yes                     & Yes                     & Yes\\  
      \midrule
      \emph{Fit statistics}\\
      Observations            & 126                     & 126                     & 126                     & 126                     & 126                     & 126\\  
      R$^2$                   & 0.88131                 & 0.89730                 & 0.84262                 & 0.91405                 & 0.93921                 & 0.88433\\  
      Within R$^2$            & 0.49208                 & 0.79015                 & 0.36957                 & 0.53003                 & 0.83278                 & 0.35453\\  
      \midrule \midrule
      \multicolumn{7}{l}{\emph{Clustered (category) standard-errors in parentheses}}\\
      \multicolumn{7}{l}{\emph{Signif. Codes: ***: 0.01, **: 0.05, *: 0.1}}\\
   \end{tabular}  
   }

\vspace{+3mm}
\caption{The Impact of Generative AI on Market Equilibrium}
\label{tab:reg_main}
\end{table*}

%% file: sections/7.demand.tex
\begin{figure*}[h]
  \centering
  \includegraphics[width=\textwidth]{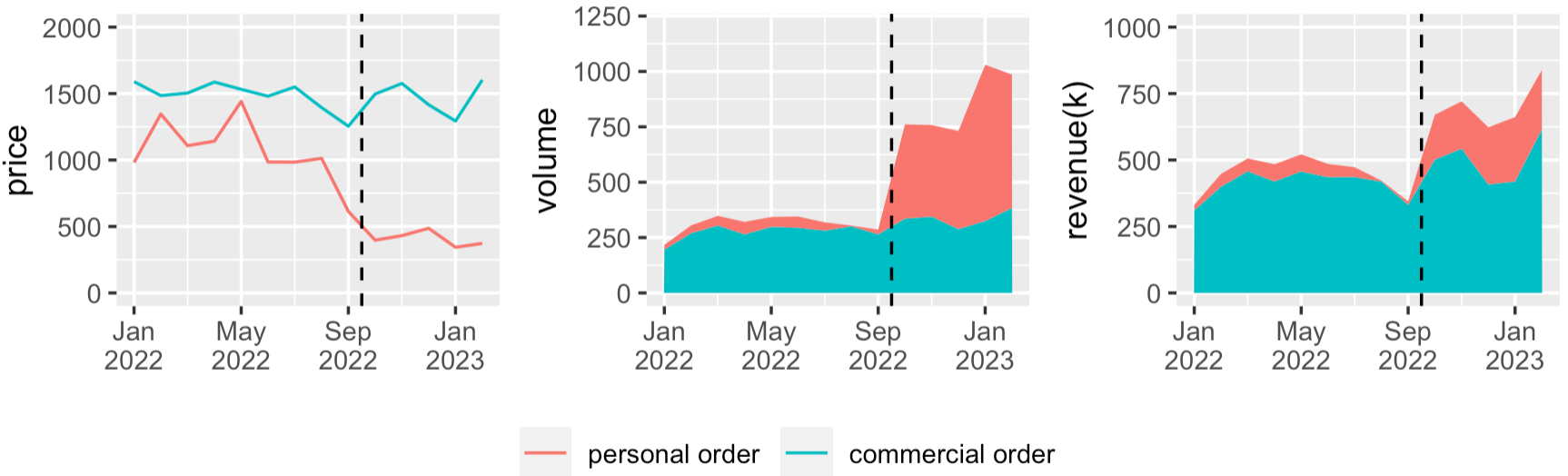}
  \caption{Market Equilibrium of Personal and Commercial Orders in the "tachie" Market}
  % \caption{Model-free evidence. The three graphs show the price, volume, and turnover trends over time for two art categories, "wallpaper" and "tachie," anime character paintings. The dashed lines in three graphs represent the leak of an anime generative AI, which is expected to affect the "tachie" market but not the "wallpaper" market. After the leak, although the price of "tachie" decreased, its volume and turnover increased significantly, suggesting that generative AI lowers price but increases volume and turnover.}
  \label{fig:personal}
\end{figure*}

\vspace{-5mm}
\section{Demand Side Analysis}
In this section, we answer research question RQ2, which aims to identify the specific demand growth that drives market prosperity\footnote{"Once you start thinking about growth, it's hard to think about anything else," said Robert Lucas, a Nobel Prize-winning economist.}. To this end, the heterogeneous effects on order licenses are examined.
% Our previous results show that the market propensity after generative AI is mainly driven by order volume growth, the source of which is interesting and meaningful\footnote{"Once you start thinking about growth, it's hard to think about anything else," said Robert Lucas, a Nobel Prize-winning economist.} to investigate. In this section, we answer research question RQ2 by examining the heterogeneous effects on order licenses.

The orders on the focal platform are classified into two categories based on licenses: personal orders and commercial orders. The artwork of personal orders is for personal use only, whereas the artwork of commercial orders is permitted for commercial purposes. In particular, commercial orders are granted more extensive property rights than personal orders, such as distribution rights and rental rights.

Figure~\ref{fig:personal} shows the market equilibrium outcomes of personal orders and commercial orders in the ”tachie” market. Prior to and following the advent of generative AI, the price, volume, and revenue of commercial orders remain relatively stable. In contrast, personal orders exhibit a substantial increase in volume and revenue with a corresponding decrease in price, which aligns with the overall "tachie" market. This suggests that the growth in the "tachie" market is primarily driven by personal orders.
% Figure~\ref{fig:personal} shows the market equilibrium outcomes of personal orders and commercial orders in the ”tachie” market. The price, volume, and revenue of commercial orders are relatively stable before and after the emergence of generative AI. Moreover, personal orders have a huge increase in volume and revenue with the decrease in price, which is consistent with the overall "tachie" market, suggesting that the growth in the "tachie" market is mainly driven by personal orders.

\input{sections/demand_table}

Formally, we replace the term, $treated$, in Equation \ref{equation:main} with $personal\_treated$ and $commercial\_treated$ to capture heterogeneous treatment effects for different order licenses:
\vspace{-3mm}
\begin{equation} \label{equation:demand}
y_{it}=\alpha_i +\delta_t + \beta_1\cdot personal\_treated_i after_t + \beta_2\cdot commercial\_treated_i after_t + \varepsilon_{it}
\end{equation}

Table~\ref{tab:reg_demand} presents the heterogeneous impact of Generative AI with different licenses. The results indicate that both personal and commercial orders exhibit a similar trend with regard to the primary average effects, namely a reduction in price and an increase in volume and revenue. Furthermore, the results demonstrate that the magnitude of the estimators for personal orders is notably higher than that of commercial orders, suggesting that generative AI benefits the "tachie" market by increasing the demand for personal orders with lower prices.
% Table~\ref{tab:reg_demand} presents the heterogeneous impact of Generative AI with different licenses. The results indicate that both personal orders and commercial orders share the same direction with the main average effects: a decrease in price and an increase in volume and revenue. What's more, our results also show that the magnitude of personal orders' estimators is remarkably higher than commercial orders' estimators, suggesting that generative AI benefits the "tachie" market by boosting the demand for personal orders with lower prices.

Our findings indicate that, as a blue ocean strategy \cite{kim2014blue}, generative AI can achieve its success by fulfilling the "low-end" needs of individual users, which were previously unmet due to the high cost. In contrast, the requisite quality for commercial orders exceeds the capabilities of current generative AI. For example, there is a notable disparity between the demo images produced by NovelAI (Figure~\ref{fig:novelai} in Appendix) and successful commercial anime images (such as \cite{HUTAO}) that feature considerably more intricate and detailed designs. Conversely, individual users may not require a very high level of quality for their orders. Generative AI can fulfill their demand directly or with a slight modification by human creators. Furthermore, generative AI may present potential copyright issues that limit its commercial applications. This is because the training of generative AI often involves the use of a substantial number of images collected from the Internet without copyright permission.
% Our findings suggest that as a blue ocean strategy \cite{kim2014blue}, generative AI can achieve its success by fulfilling the "low-end" needs of individual users, that were not fulfilled before because of expensive cost. In contrast, commercial orders require a high quality, which is beyond the ability of the current generative AI. For example, there is a remarkable gap between Novelai's demo images (Figure~\ref{fig:novelai} in Appendix) and successful commercial anime images (such as \cite{HUTAO}) with much more delicate and detailed designs. On the other hand, personal users may not require a very high quality for their orders. Generative AI can fulfill their demand directly or with a slight modification by human creators. In addition, generative AI has potential copyright problems which limits commercial uses, because training generative AI uses a bulk of images collected from the Internet without copyright permission.

% One possible reason is that generative AI suffers from copyright problems because training generative AI uses many images collected from the Internet without copyright permission, which is hard for commercial purposes. Another possible reason is that the quality of generative AI images is not enough for commercial uses. For example, there is a remarkable gap between Novelai's demo images and successful commercial anime images with much more delicate and detailed designs. On the other hand, personal users may not require a very high quality for their orders. Generative AI can fulfill their demand directly or with a slight modification by artists. 

%% file: sections/demand_table.tex
\begin{table*}[h!]
% \linespread{1}
% \huge
% \selectfont
% \footnotesize
\centering
    % \vspace{-5mm}
    \resizebox{\linewidth}{!}{
   \begin{tabular}{lcccccc}
      \tabularnewline \midrule \midrule
     Dependent Variables:                 & price                    & volume                   & revenue                 & log(price)               & log(volume)              & log(revenue)\\  
      Model:                               & (1)                      & (2)                      & (3)                      & (4)                      & (5)                      & (6)\\  
      \midrule
      \emph{Variables}\\
      personal\_treated $\times$ after     & -746.0$^{**}$            & 129.5$^{***}$            & 41,244.4$^{***}$         & -0.9892$^{***}$          & 2.953$^{***}$            & 1.964$^{***}$\\   
                                           & (17.61)                  & (1.520)                  & (195.6)                  & (0.0101)                 & (0.0162)                 & (0.0061)\\   
      commercial\_treated $\times$ after   & -50.57$^{***}$           & 37.25$^{***}$            & 21,115.6$^{***}$         & -0.0967$^{***}$          & 0.3903$^{***}$           & 0.2936$^{***}$\\   
                                           & ($1.36\times 10^{-12}$)  & ($2.62\times 10^{-13}$)  & ($1.03\times 10^{-10}$)  & ($2.27\times 10^{-15}$)  & ($6.77\times 10^{-15}$)  & ($4.43\times 10^{-15}$)\\    
      \midrule
      \emph{Fixed-effects}\\
      art\_category                        & Yes                      & Yes                      & Yes                      & Yes                      & Yes                      & Yes\\  
      year\_week                           & Yes                      & Yes                      & Yes                      & Yes                      & Yes                      & Yes\\  
      \midrule
      \emph{Fit statistics}\\
      Observations                         & 185                      & 185                      & 185                      & 185                      & 185                      & 185\\  
      R$^2$                                & 0.78796                  & 0.91291                  & 0.82473                  & 0.79054                  & 0.92775                  & 0.83418\\  
      Within R$^2$                         & 0.59778                  & 0.73866                  & 0.77730                  & 0.63074                  & 0.85464                  & 0.76433\\  
      \midrule \midrule
      \multicolumn{7}{l}{\emph{Clustered (art\_category) standard-errors in parentheses}}\\
      \multicolumn{7}{l}{\emph{Signif. Codes: ***: 0.01, **: 0.05, *: 0.1}}\\

   \end{tabular}  
   }

\vspace{+3mm}
\caption{The Impact of Generative AI on Market Equilibrium - Heterogeneity by Licenses}
\label{tab:reg_demand}
\end{table*}

%% file: sections/8.supply.tex
\input{sections/supply_table}

\section{Supply Side Analysis}
Our previous findings indicate that generative AI has the potential to benefit artists by increasing their order volume and revenue. However, it remains unclear whether incumbents or new entrants utilizing generative AI will reap most of the benefits. In this section, we answer research question RQ3, by examining the incumbents in the market.
% Our previous results show that generative AI benefits artists by increasing their order volume and revenue. However, it remains unclear whether incumbents or new entrants using generative AI will reap most of the benefits. In this section, we answer research question RQ3, by examining the incumbents in the market.

We focus on the completed orders in order to observe which artists won the orders. We define incumbents as the group of artists who registered prior to the leak, i.e., before October 2022. We will examine the market equilibrium before and after the leak with the orders completed by incumbents.
% We limit our attention to the completed orders so that we can observe which artists won the orders. We define incumbents as the group of artists who registered before the leak, i.e., before October 2022. We will examine the market equilibrium before and after the leak with the orders completed by incumbents.

We employ the standard difference-in-differences in Equation (1). Table~\ref{tab:reg_supply} presents the estimation results. Panel A's dataset comprises all completed orders, whereas Panel B's dataset comprises only orders completed by incumbents. The results in Panel B are consistent with our main effects, indicating that incumbents benefit from generative AI by gaining more orders (116\%) and more revenue (65\%). Furthermore, the estimators in Panel B is close to those in Panel A, indicating that the majority of the orders were completed by incumbents, resulting in a similar sample for Panel A and B. According to our data, the incumbents won 97.41\% of the orders and 97.96\% of the revenue after the leak.
% We continue using the standard difference-in-differences in Equation (1). Table~\ref{tab:reg_supply} presents the estimation results. Panel A's data uses all the completed orders and panel B's data uses the orders completed by the incumbents. We find that the results in Panel B are consistent with our main effects, showing that incumbents benefit from generative AI by gaining more orders (116\%) and more revenue (65\%). In addition, the estimators in Panel B are very close to the estimators in Panel A, indicating that most of the orders were completed by incumbents, constructing a similar sample for Panel A and B. According to our data, the incumbents won 97.41\% of the orders and 97.96\% of the revenue after the leak.

One potential explanation for this result is that the current generative AI is unable to fulfill orders without human artists. This is because the generation of content is an iterative process between consumers and artists, with artists required to make ongoing modifications in response to consumer feedback. The current generative AI, which outputs images, audio, and videos, lacks the capacity to make slight modifications independently due to the limitations of its underlying algorithms.
% One possible reason for this finding is that generative AI alone cannot fulfill orders without painting expertise because it is an iterative process between consumers and artists. Artists have to continue to make modifications according to the requirements of consumers. The current generative AI that takes prompts as input and outputs images, audio, and videos cannot make slight modifications alone because of the limitations of their inherent algorithms.

Our results rule out the possibility that new entrants using generative AI have taken over the market share of genuine human creators. In contrast, generative AI serves to expand the market itself, with the incumbent creators deriving benefits from such expansion. The findings provide empirical evidence of a favorable economic consequence of generative AI for artists and suggest that they should embrace generative AI rather than resist it. The incorporation of generative AI into one's workflow has the potential to increase productivity and leverage the novelty and enhanced capabilities provided by generative AI to attract new customers who previously were not engaging with the art market, thereby "growing the pie."
% Our results rule out the possibility that new entrants using generative AI have taken over the market share of genuine human creators. In contrast, generative AI expands the market itself greatly and the incumbents reap the benefits of such expansion. The findings provide empirical evidence of a favorable economic consequence of generative AI for artists and suggest that they should welcome generative AI instead of resisting it. They should incorporate generative AI into their workflow to increase their productivity and leverage the novelty and enhanced capabilities provided by generative AI to attract new customers who previously were not engaging with the art market to "grow the pie".

%% file: sections/supply_table.tex
\begin{table*}[!t]
% \linespread{1}
% \huge
% \selectfont
% \footnotesize
\centering
    % \vspace{-5mm}
    \resizebox{\linewidth}{!}{
   \begin{tabular}{l|ccc|ccc}
      % \tabularnewline \midrule \midrule
      \midrule \midrule
       & \multicolumn{3}{c|}{Panel A: All Completed Orders}                         & \multicolumn{3}{c}{Panel B: Incumbent Completed Orders} \\
      Dependent Variables:    & log(price)               & log(volume)              & log(revenue)            & log(price)              & log(volume)             & log(revenue)\\  
      Model:                  & (1)                      & (2)                      & (3)                      & (4)                     & (5)                     & (6)\\  
      \hline

      \emph{Variables} & & & & & &\\
      treated $\times$ after  & -0.4915$^{***}$          & 1.171$^{***}$            & 0.6790$^{***}$           & -0.5127$^{***}$         & 1.168$^{***}$           & 0.6556$^{***}$\\   
                              & ($1.33\times 10^{-15}$)  & ($3.51\times 10^{-15}$)  & ($2.02\times 10^{-15}$)  & ($1.5\times 10^{-15}$)  & ($3.1\times 10^{-15}$)  & ($1.92\times 10^{-15}$)\\    
      \hline
      \emph{Fixed-effects} & & & & & &\\
      art\_category           & Yes                      & Yes                      & Yes                      & Yes                     & Yes                     & Yes\\  
      year\_week              & Yes                      & Yes                      & Yes                      & Yes                     & Yes                     & Yes\\  
      \hline
      \emph{Fit statistics} & & & & & &\\
      Observations            & 126                      & 126                      & 126                      & 126                     & 126                     & 126\\  
      R$^2$                   & 0.79706                  & 0.89378                  & 0.73652                  & 0.77257                 & 0.89462                 & 0.73014\\  
      Within R$^2$            & 0.21484                  & 0.67848                  & 0.19868                  & 0.20654                 & 0.67798                 & 0.17595\\  
      \midrule \midrule
      \multicolumn{7}{l}{\emph{Clustered (art\_category) standard-errors in parentheses}}\\
      \multicolumn{7}{l}{\emph{Signif. Codes: ***: 0.01, **: 0.05, *: 0.1}}\\

   \end{tabular}  
   }

\vspace{+3mm}
\caption{The Impact of Generative AI on Market Equilibrium with Orders Completed by Incumbents}
\label{tab:reg_supply}
\end{table*}

%% file: sections/8.conclusions.tex
\section{Conclusions}
This paper examines the impact of the advent of generative AI on market equilibrium, in terms of price, volume, and revenue. We identify an unanticipated and sudden leak of an advanced image-generative AI as a natural experiment, which facilitates difference-in-differences comparisons for causal inference. The results indicate that generative AI substantially increases order volume and overall revenue with a corresponding decrease in average price. The increase in volume and revenue is primarily driven by the rising demand for personal orders with lower prices, rather than expensive commercial orders. Moreover, the incumbent creators maintained the majority of the market share, reaping most of the benefits from generative AI. 
% The results are explained theoretically through the lens of market equilibrium theory.

This paper makes a significant scholarly contribution to the growing literature on the economics of AI, by providing one of the first empirical evidence on how generative AI affects the platform economy through diverse perspectives. Moreover, the findings also offer valuable practical insights. Creators should embrace generative AI rather than resist it to gain greater benefits from enhanced productivity. Online platforms can conduct marketing campaigns and improve services specifically for "low-end" personal orders, the emerging demand driven by generative AI. Policymakers may also implement supportive policies for generative AI to create more jobs and tax revenue.

It is our contention that generative AI is having and will continue to have a profound impact on our world. In the context of the marketplace, future research can explore how generative AI affects product quality, consumer satisfaction, and beyond through our identified natural experiment. 

\textbf{Acknowledgments.} We would like to thank Reviewer HvRd, QbGb, and Area Chair Y8tc for their constructive feedback. This work was supported by the National Natural Science Foundation of China (Grant No.92370204), National Key R\&D Program of China (Grant No.2023YFF0725001), Guangzhou-HKUST(GZ) Joint Funding Program (Grant No.2023A03J0008), Education Bureau of Guangzhou Municipality.

% This paper makes a significant contribution to the growing literature on the economics of AI, by providing one of the first empirical evidence on how generative AI affects the platform economy through diverse perspectives. Furthermore, the findings offer valuable practical insights for creators, platform owners, and policymakers regarding the exploitation and management of generative AI to enhance its utility.

% This paper provides empirical evidence that generative AI, while decreasing price, surprisingly increases volume and revenue in product markets, as a result of decreasing demand and increasing supply.

% We identified an unexpected and sudden leakage of a highly proficient image-generating AI as a natural experiment for causality. This natural experiment allows for answering many critical research questions. A new research question could delve deeper into the outcome variables in this paper, such as how demand/orders change after generative AI. Also, a new research question could be another important outcome variable, such as what's the impact of generative AI on product quality. In addition, a new research question could relate to another area, such as what's the impact of generative AI on labor markets. 